\documentclass[journal]{IEEEtran}
\usepackage{xcolor}
\usepackage{graphicx}
\usepackage{float}
\usepackage{siunitx}
\usepackage{array}
\usepackage[flushleft]{threeparttable}
\usepackage[utf8]{inputenc}

\begin{document}
\pagenumbering{arabic}
\title{A 0.5GHz 0.35mW LDO-Powered Constant-Slope Phase Interpolator with 0.22$\%$ INL}

\author{\IEEEauthorblockN{Ahmed Elnaqib,
Hayate Okuhara, 
Taekwang Jang,
Davide Rossi and 
Luca Benini}}

\maketitle

\begin{abstract}
Clock generators are an essential and critical building block of any communication link, whether it be wired or wireless, and they are increasingly critical given the push for lower I/O power and higher bandwidth in Systems-on-Chip (SoCs) for the Internet-of-Things (IoT). One recurrent issue with clock generators is multiple-phase generation, especially for low-power applications; several methods of phase generation have been proposed, one of which is phase interpolation. We propose a phase interpolator (PI) that employs the concept of constant-slope operation. Consequently, a low-power highly-linear operation is coupled with the wide dynamic range (i.e. phase wrapping) capabilities of a PI. Furthermore, the PI is powered by a low-dropout regulator (LDO) supporting fast transient operation. Implemented in 65-nm CMOS technology, it consumes 350$\mu$W at a 1.2-V supply and a 0.5-GHz clock; it achieves energy efficiency 4$\times$-15$\times$ lower than state-of-the-art (SoA) digital-to-time converters (DTCs) and an integral non-linearity (INL) of 2.5$\times$-3.1$\times$ better than SoA PIs, striking a good balance between linearity and energy efficiency.
\end{abstract}


\blfootnote{The authors are with the University of Bologna, Bologna, Italy and ETH Zurich, Zurich, Switzerland (e-mail: ahmedgamal.mahmoud2@unibo.it, hayate.okuhara@unibo.it, tkjang@ethz.ch, davide.rossi@unibo.it, luca.benini@u\-nibo.it)}

\IEEEpeerreviewmaketitle

\section{Introduction}
\begin{figure}[t]
     \centering
     \includegraphics[width=3.4in]{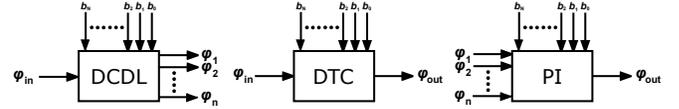}
     \caption{Types of digital-to-phase converters (DPCs).}
     \label{dpcs}
 \end{figure}

 Due to the importance of multiple-phase generation, several methods have been presented tackling the main trade-offs that constrain phase generators: resolution, dynamic range, linearity, and power consumption. Different applications have led to different solutions dictated by their specific requirements. Phase generators, or digital-to-phase converters (DPCs) as we will be strictly concerned with digital control, can be classified into digitally-controlled delay lines (DCDLs), digital-to-time converters (DTCs) and phase interpolators (PIs) (Fig. \ref{dpcs}).

Starting with DCDLs, a delay line is a single-input multiple-output DPC based on identical delay cells. The minimum resolution of a DCDL is limited by the minimum delay of the delay cell, which in turn is limited by the minimum inverter delay. The minimum operating frequency is limited by the maximum delay of the delay line, which is limited by the number of delay stages used, and more stages mean higher power consumption \cite{dll}. Therefore, there is an inherent trade-off between resolution, dynamic range, and power. Furthermore, a delay line could be followed by a multiplexer (MUX) which implies that the MUX delay has to be taken into account as well \cite{dcdlnand}. Delay lines are also susceptible to PVT variations, which compromises linearity. Therefore, DCDLs are typically used within a delay-locked loop (DLL) in order to control these mismatches, resulting in extra overhead \cite{dcdlmod}, \cite{dcdlmirror}.
 
 \begin{figure}
    \centering
    \includegraphics[width=2.5in]{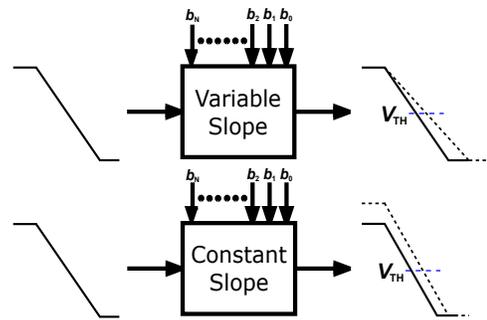}
    \caption{Constant-slope vs. variable-slope operation.}
    \label{constvsvar}
    \vspace{-5mm}
\end{figure}

\begin{table*}[t]
\caption{State-of-the-Art DTCs Comparison}
\label{dtcs}
\centering
\begin{tabular}{ | c | c | c | c | c | c | c | c |} 
\hline
& \cite{dtc1} & \cite{dtc2} & \cite{dtc4}
& \cite{dtc_load} & \cite{dtc_dcdl} & \cite{dtc_constant} & \cite{dtc_const2}\\
\hline
\hline
Technology & 28-nm & 40-nm & 65-nm & 28-nm & 40-nm & 65-nm & 65-nm\\
\hline
Architecture & Variable-slope & Variable-slope & Multi-stage Variable-slope & Variable-slope & Delay Line & Constant-slope & Constant-slope\\
\hline
Supply Voltage & 0.9V & 1.1V & 0.9V & 0.9V & 1.1V & 1.2V & 1.0V\\
\hline
Frequency & 40MHz & 150MHz & 2GHz & 40MHz & 50MHz & 55MHz & 52MHz\\
\hline
Dynamic Range & 512ps & 1.1ns & 256ps & 563ps & 1.28ns & 189ps &  593ps \\ 
\hline
Resolution & 500fs & 1.075ps & 2ps & 550fs & 16ps & 185fs & 580fs \\ 
\hline
INL & 1.5ps (0.29$\%$) & 2.8ps (0.25$\%$) & 2ps (0.78$\%$) & 990fs (0.18$\%$) & 1.4ps (0.11$\%$) & 328fs (0.17$\%$) & 870fs (0.15$\%$) \\ 
\hline
Power & 0.5mW & - & - & 0.58mW & 0.53mW & 0.8mW & 0.14mW\\
\hline
\end{tabular}
\vspace{-5mm}
\end{table*}

 Moving on to the DTC, it is a single-input single-output block that typically generates delay through a comparator that detects the threshold crossing of a current charging a capacitor. Some delay cells that are used in delay lines can actually be considered stand-alone DTCs (e.g. current-starved inverter). DTCs are widely used in fractional-N frequency synthesizers for quantization noise cancellation \cite{dtc1}-\cite{dtc3}; DTCs provide fine resolution, but it can be difficult to achieve a wide dynamic range while maintaining an acceptable linearity \cite{pi_pipe}. In \cite{dtc_load}, the DTC is composed of a delay stage loaded by switched capacitors; it can be classified as a variable-slope DTC and will be discussed later in more detail. A delay-line-based DTC can be found in \cite{dtc_dcdl} which is basically a delay line with a single output, but the input is injected into any given delay cell based on a digital control code. The delay cells are identical inverters with a slight mismatch between even and odd cells. While it consumes as much as 0.53mW at 50MHz, it achieves 0.11$\%$ INL with constant delay cell currents.

\begin{figure}[t]
    \centering
    \includegraphics[width=1.5in]{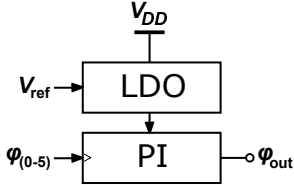}
    \caption{Simplified block diagram of the proposed DPC.}
    \label{dpc}
    \vspace{-5mm}
\end{figure}

 Another approach is the constant-slope DTC \cite{dtc_constant}. Most DTCs depend upon the delay generated by a voltage ramp \cite{dtc1}-\cite{dtc_load} and this delay is mainly adjusted through a tunable capacitance; also, some DCDLs use switched current sources as well \cite{dcdlmirror}; both of these methods tune the delay by varying the slope of the output signal. This was shown to be a source of non-linearity due to the relationship between the input slope and the delay of the comparator used to detect these signals; the bandwidth limit of the comparator could be another source of non-linearity as well \cite{dtc_constant}. Therefore, a constant slope was proposed by varying the start voltage rather than the load capacitance as shown in Fig. \ref{constvsvar}. In \cite{dtc_const2}, a DTC based on this concept was proposed for low-power applications achieving an INL of 0.15$\%$. Shown in Table \ref{dtcs} is a comparison of the different DTC implementations that have been discussed. It can be seen that constant-slope DTCs exhibit better linearity compared to variable-slope DTCs.

PIs exploit multiple clock phases in order to generate an intermediate phase. PIs can be found in fractional feedback dividers \cite{pi_current}, \cite{pi_pipe} and clock/data recovery (CDR) systems \cite{pi_current2}. In \cite{pi_current}, the tail currents are manipulated to adjust the weight of each input phase. The varying slope of the tail current is a major source of non-linearity in such current-mode interpolators \cite{pi_pipe}. They also suffer from the static power consumption typical of CML-based circuitry. A charge-mode PI was proposed in \cite{me} which makes use of a tunable capacitance to produce different phases similar to DTCs based on a voltage ramp. However, to improve linearity, non-linear capacitance values were used to counteract the variable slope operation; this makes calibration and matching extremely challenging, and leads to a degradation in linearity. Another issue is the limited support for multi-rate operation because it requires tuning a full array of capacitors, rather than a single capacitance. Alternatively, a pipelined PI is proposed in \cite{pi_pipe} that consists of several stages, each capable of phase forwarding or phase interpolation. This scheme aims at improving the linearity by maintaining a constant current flowing through each stage, which is the same concept proposed by constant-slope DTCs. This indicates a trend of adopting constant-slope operation in both DTCs and PIs.

\begin{figure}[t]
    \centering
    \includegraphics[width=1.72in]{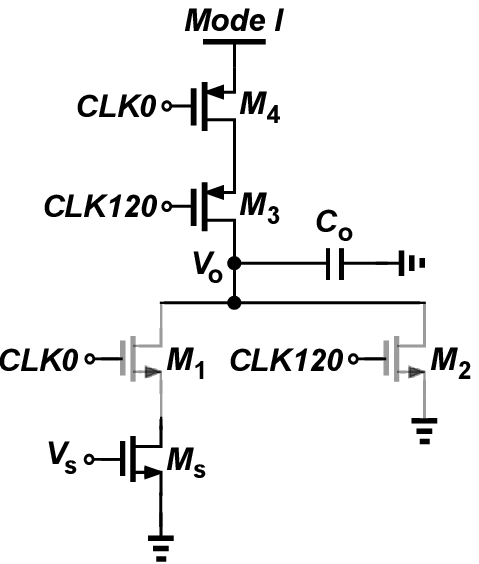}
    \includegraphics[width=1.72in]{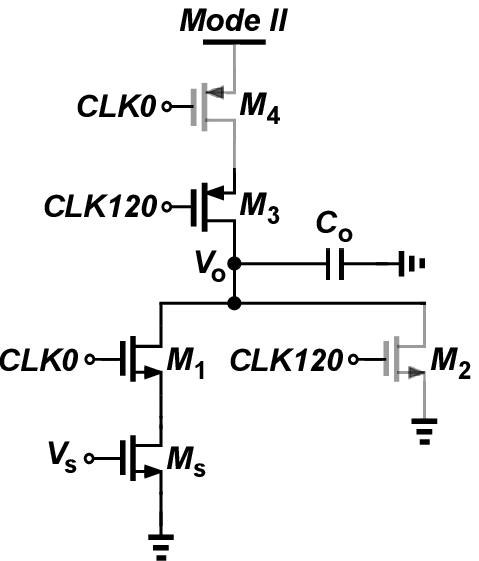}
    \includegraphics[width=3.2in]{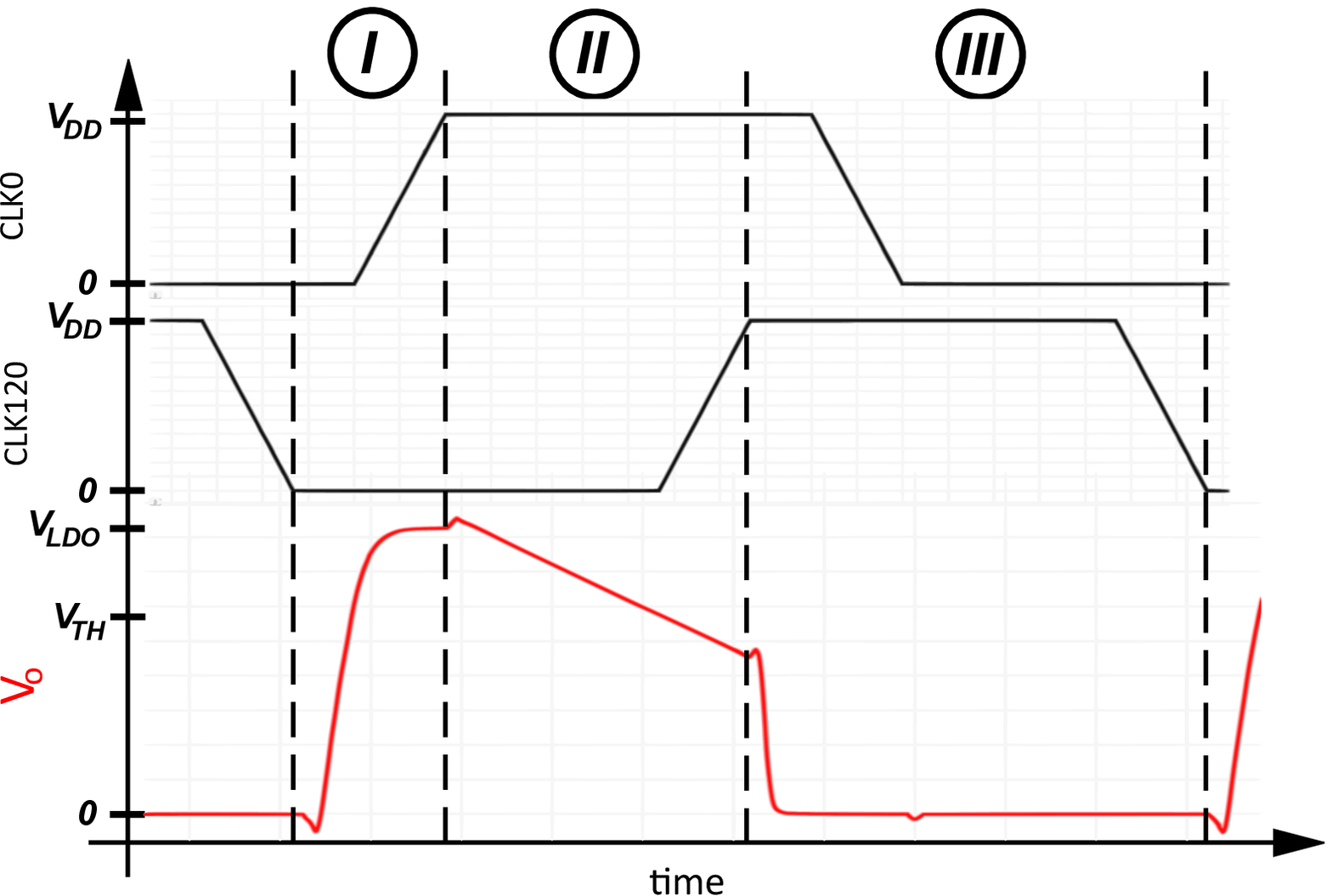}
    \caption{Schematic and operation of a PI half-cell.}
    \label{pi_op}
    \vspace{-5mm}
\end{figure}

In this paper, we present a highly-linear DPC suited to low-power applications that demonstrates the similarities between state-of-the-art PIs and DTCs, and draws inspiration from both; it achieves the wide dynamic range of a PI and the linearity of constant-slope signaling. It improves the energy efficiency of DTCs by 4$\times$-15$\times$ and keeps the INL at 2$\times$ at worst; and when compared to PIs, linearity is improved by 2.5$\times$-3.1$\times$ while maintaining the energy efficiency within 1.1$\times$-1.5$\times$. In Section II, the proposed DPC design is described in detail, and in Section III, the simulation results of the DPC are presented and discussed. Section IV is the conclusion.

\section{Proposed Phase Generator Operation}

Shown in Fig. \ref{dpc} is a simplified block diagram of the proposed DPC. The PI exploits constant-slope operation, so a low-dropout regulator (LDO) provides the supply voltage (i.e. start voltage) corresponding to a given output phase. The PI was implemented within a CDR loop; the loop has an update period of 128ns and is clocked by six phases provided by an external frequency-locked loop (FLL) \cite{fll} and integrated with a low-power micro-controller system. The phase space is divided into six regions (sextants) according to the six-phase clock provided by the FLL; therefore, six PI unit cells are needed.

\subsection{Proposed Phase Interpolator}

\begin{figure} [t]
    \centering
    \includegraphics[width=3.2in]{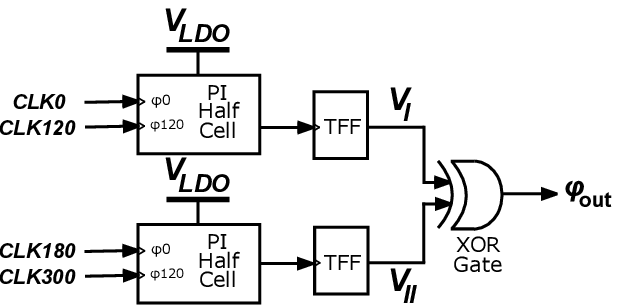}
    \includegraphics[width=3.2in]{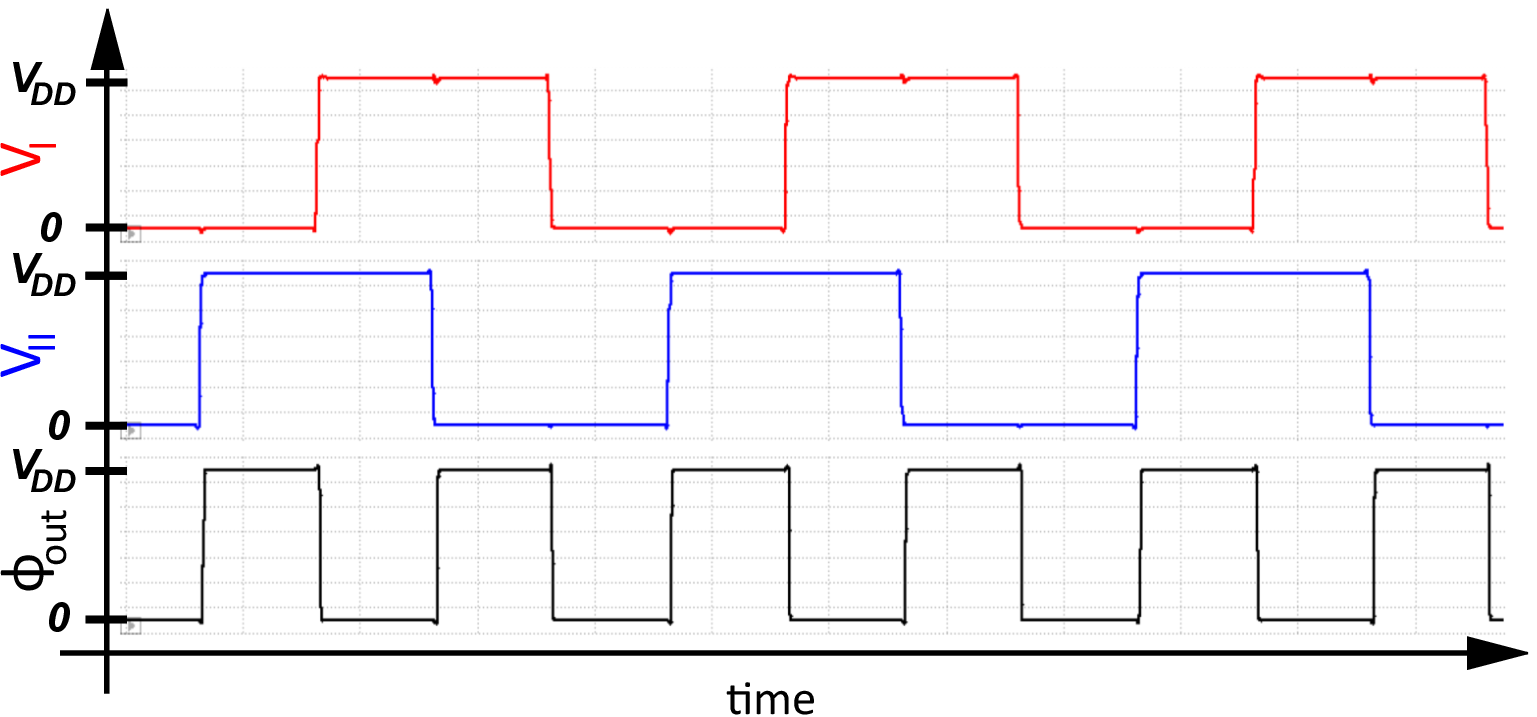}
    \caption{Phase interpolator unit cell.}
    \label{picell}
\end{figure}

Shown in Fig. \ref{pi_op} is the schematic of a PI half-cell. $M_s$ is a constant current source and $C_o$ is the output capacitance. $M_1$ controls the discharge timing of $C_o$; $M_2$ is a reset switch to discharge $C_o$ to ground; $M_3$ and $M_4$ are set switches that charge $C_o$ to $V_{LDO}$. The operation of the half-cell is shown as well. In Mode I, where CLK0 and CLK120 are low, $C_o$ is charged from 0 to $V_{LDO}$ (i.e. start voltage); in Mode II, where only CLK0 is high, $C_o$ discharges through $M_1$, and it is during this mode that the threshold crossing is detected by subsequent comparators. Then, $C_O$ continues to discharge through $M_2$ as well when CLK120 is high, and does so exclusively when CLK0 eventually goes low; that is Mode III, which is the reset mode.

\begin{figure}[t]
    \centering
    \includegraphics[width=3.4in]{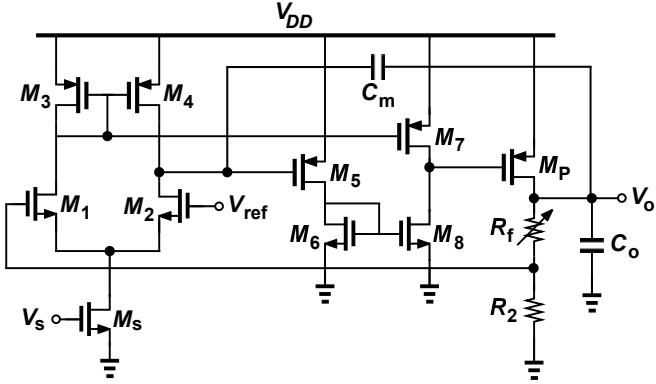}
    \caption{Low-dropout regulator (LDO) schematic.}
    \label{LDO}
    \vspace{-5mm}
\end{figure}

The slope of the output voltage ($V_o$) during the relevant Mode II can be represented as \[\frac{dV_o}{dt} = -\frac{I_c}{C_o}\] where $I_c$ is the current flowing through $M_1$. Since $C_o$ is constant, then in order to achieve constant-slope operation, $I_c$ needs to be kept constant across the different PI steps. To this end, a constant current source $M_s$ is used. Therefore, the slope of the output voltage can be solely controlled by $C_o$, which can be tuned to support multi-rate operation. The maximum current variation across the PI steps is kept at about 5$\%$.

To build a unit cell, the output of two PI half-cells are then connected to toggle flip-flops (TFFs) as shown in Fig. \ref{picell}; the inputs of the second half-cell are the inverse of the inputs to the first half-cell. The TFFs provide two half-rate clock signals that are strictly $90^o$ apart since they are generated by two full-rate clock signals that are $180^o$ apart. Finally, feeding both to a 2-input XOR gate generates the required full-rate clock signal. Accordingly, the output clock phase is controlled by the input clock phases (i.e. the phases fed to a given unit cell), and the start voltage provided by the LDO.

\begin{figure}[t]
    \centering
    \includegraphics[width=3.2in]{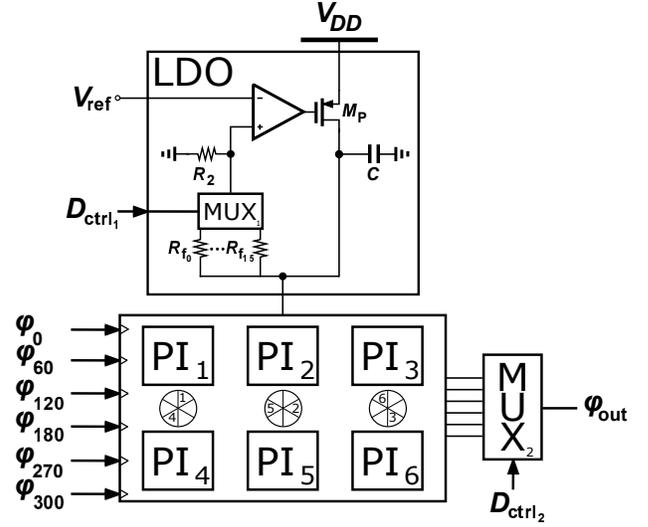}
    \caption{Block diagram of the full proposed DPC.}
    \label{fulldpc}
\end{figure}

\begin{figure}[t!]
    \centering
    \includegraphics[width=2.8in]{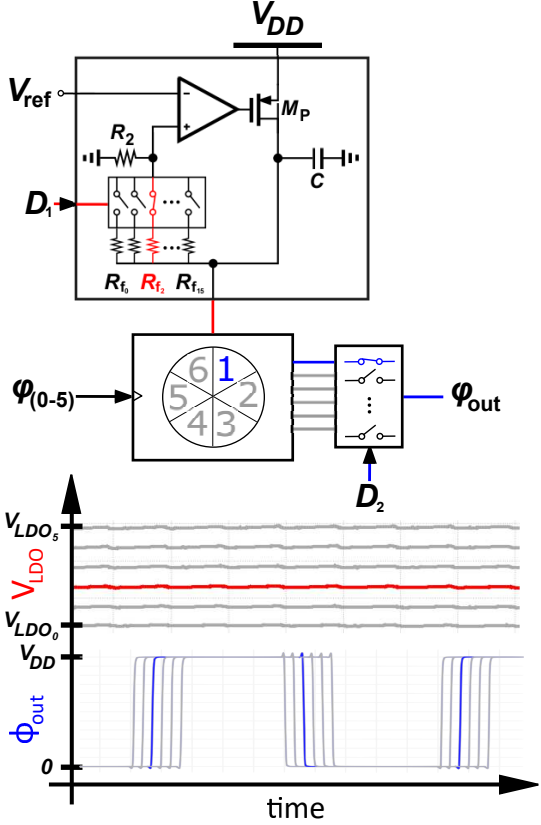}
    \caption{Proposed DPC operation with an output phase of 22.5$^o$. (sextant 1, step 3)}
    \label{dpcex}
\end{figure}

\begin{figure}[t]
    \centering
    \includegraphics[width=3.4in]{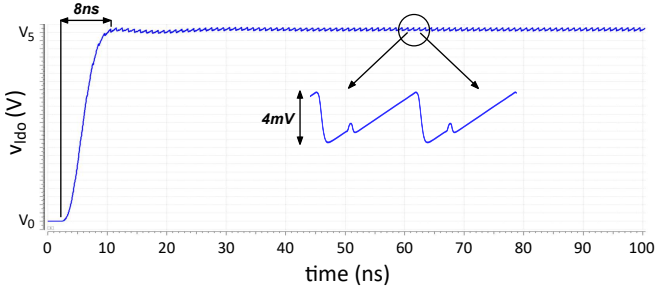}
    \caption{LDO step response.}
    \label{LDOstep}
\end{figure}

\begin{figure}[t]
    \centering
    \includegraphics[width=3.4in]{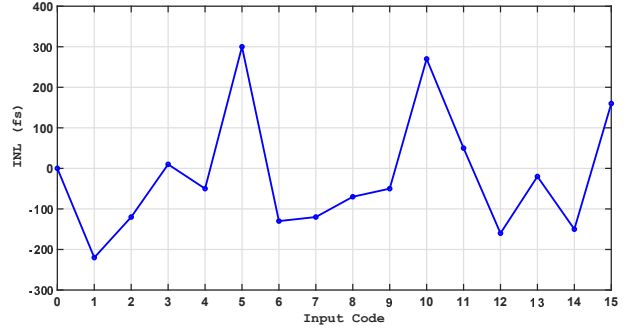}
    \caption{INL Nominal Simulation.}
    \label{INL}
\end{figure}

\begin{figure}[t]
    \centering
    \includegraphics[width=3.6in]{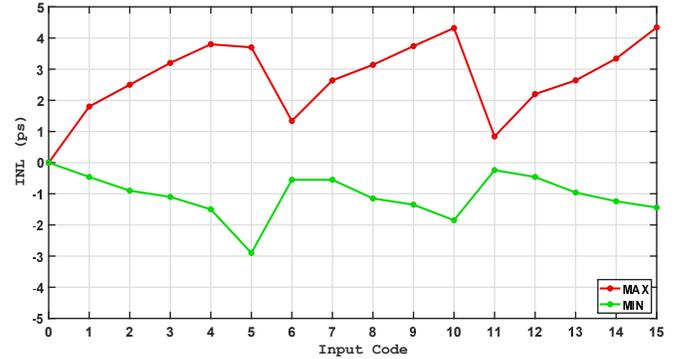}
    \caption{INL Monte Carlo Simulation.}
    \label{INLc}
\end{figure}

\begin{figure}[t]
    \centering
    \includegraphics[width=3.1in]{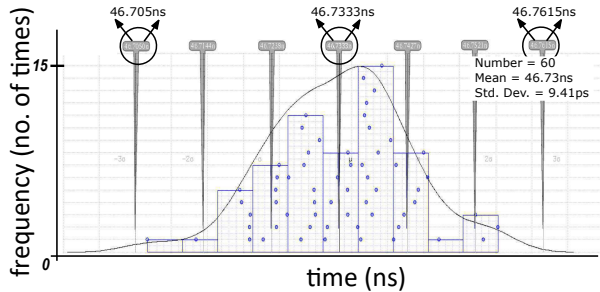}
    \caption{Zero-crossing histogram of output phase 22.5$^o$.}
    \label{hist}
\end{figure}

\begin{figure}[t]
    \centering
    \includegraphics[width=3.4in]{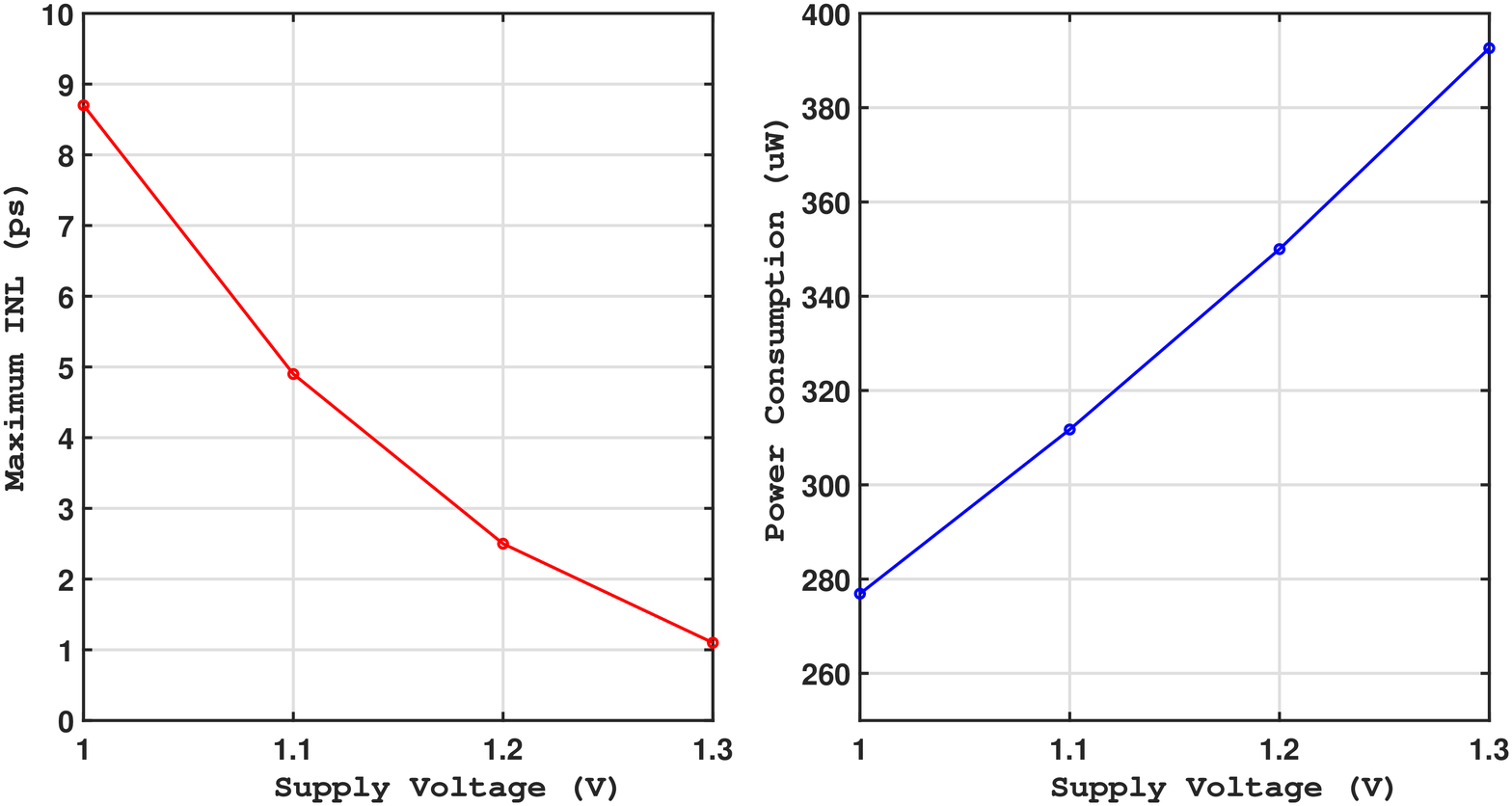}
    \caption{MC INL and power consumption versus supply voltage for output phase 22.5$^o$.}
    \label{inl_v_supply}
\end{figure}

\subsection{Low-Dropout Regulator}
Shown in Fig. \ref{LDO} is the schematic of the LDO. Since high gain and fast response are the main targets, a three-stage regulator is used; the error amplifier consists of two stages. $M_s$, $M_1$, $M_2$, $M_3$, and $M_4$ compose the first stage; the second stage consists of  $M_5$, $M_6$, $M_7$, and $M_8$; $M_P$ is the power transistor, and $R_f$ and $R_2$ provide the feedback signal; $C_m$ is the compensation capacitor and $C_o$ is the output capacitance. $V_o$ is controlled by the variable resistance $R_f$ where its value is determined by a digital control code based on the required output phase. The range of $V_o$ is from 0.7V to 1V.

The full DPC architecture is shown in Fig. \ref{fulldpc}; the control bits for the LDO and the PI are both generated by the CDR loop; the phase space is divided into sextants with the first sextant consisting of 6 steps, the second and third of 5 steps each; these three sextants cover half the phase space, and they are mirrored in the other half. The first multiplexer selects the feedback resistor for a given start voltage (i.e. $V_{LDO}$), thus handling fine control within a sextant. The second multiplexer handles coarse control by selecting the sextant. Shown in Fig. \ref{dpcex} is an example of the operation of the DPC where an output phase of 22.5$^o$ is required; since 22.5$^o$ lies in the first sextant (i.e. from 0$^o$ to 60$^o$), the first PI is selected by MUX2 (i.e. coarse control); a phase resolution of 11.25$^o$ is used, so MUX1 selects the third feedback resistor (i.e. voltage step), given that the first step is 0$^o$.

\section{Simulation Results}
The LDO has an open-loop gain of 50dB across the required range of start voltages with a phase margin of 60$^o$. The step response of the LDO is shown in Fig. \ref{LDOstep}; the largest possible output voltage transition (i.e. feedback resistor changing from $R_{f_0}$ to $R_{f_5}$) was simulated and the output voltage was observed; the LDO load is a PI unit cell. The output voltage stabilizes within 8ns; this was deemed acceptable as the sampling period of the CDR is 16ns. The ripples on the output voltage are within 4mV. The quiescent current is around 50$\mu$A.

The INL of the 5-bit PI is demonstrated in Fig. \ref{INL}; the error is within 300fs (0.015$\%$). The Monte Carlo process variations is also shown in Fig. \ref{INLc}. The worst maximum of the INL is 4.32ps (0.22$\%$), and the worst minimum INL is -2.9ps (0.15$\%$). $V_{th_n}$ is 392.73mV for the maximum points and 390.28mV for the minimum points. To combat these variations, the PI output capacitance $C_o$ can be dynamically calibrated. Only one PI unit cell is kept powered-up at a given time; a PI unit cell consumes about 290$\mu$W. 

To demonstrate the variation in the edges of the output clock, the histogram of the zero-crossings of an output phase of 22.5$^o$ is shown in Fig. \ref{hist}; the mean is around 46.73ns in this example and the standard deviation is 9.41ps. Also, shown in Fig. \ref{inl_v_supply} are the results of the Monte Carlo simulations for the maximum INL and corresponding power consumption across different values of the supply voltage. Under temperature variations, the worst INL is 3.54ps at a temperature of -40$^o$ and -0.63ps at 125$^o$, resulting in a variation of 4.17ps across this range.

\begin{table*}[t!]
\caption{State-of-the-Art DTCs and PIs Comparison}
\label{pis}
\centering
\begin{threeparttable}
\begin{tabular}{| c | c | c | c | c | c |} 
\hline
& \cite{dtc_dcdl} & \cite{dtc_const2} & \cite{pi_current2} & \cite{pi_pipe}
& \textbf{This Work} \\
\hline
\hline
Technology & 40-nm & 65-nm & 65-nm & 65-nm & 65-nm\\
\hline
Architecture & Delay line & - & Current-mode & Pipeline & Charge-mode \\
\hline
Slope & Constant & Constant & Variable & Constant & Constant \\
\hline
Supply Voltage & 1.1V & 1.0V & 1.2V & 1V & 1.2V \\
\hline
Frequency & 50MHz & 52MHz & 6GHz & 5GHz & 0.5GHz \\
\hline
Dynamic Range & 1.28ns & 593ps & 166.66ps & 200ps & 2ns\tnote{*}  \\ 
\hline
Resolution &  16ps & 580fs & 4.63ps & 6.25ps & 62.5ps\tnote{*}  \\ 
\hline
INL & 1.4ps (0.11$\%$) & 870fs (0.15$\%$) & 920fs (0.55$\%$) & 1.4ps (0.7$\%$) & 4.32ps\tnote{*} (0.22$\%$)  \\ 
\hline
Power & 0.53mW & 0.14mW & $<$3.8mW & 2.3mW & 0.35mW\tnote{*} \\
\hline
\end{tabular}
\begin{tablenotes}
\item[*] Simulation results
\end{tablenotes}
\end{threeparttable}
\end{table*}

The performance of the proposed PI is compared to other state-of-the-art PIs and DTCs in Table \ref{pis}; the DTCs with the highest linearity and the lowest power consumption from Table \ref{dtcs} were included in this comparison. For the PIs, the traditional phase deviation metric was replaced by the equivalent integral non-linearity (INL) metric to have a basis for comparing them to DTCs. The combination of a 0.7pJ/bit energy efficiency and 0.22$\%$ INL achieved by this work balances that of state-of-the-art PIs and DTCs. 

\section{Conclusion}
A constant-slope phase generator in 65-nm was presented that achieves a balance between the speed and energy efficiency of phase interpolators, and the linearity of digital-to-time converters. It employs a PI powered by a low-dropout regulator, and it has a worst peak of 4.32ps INL operating at 0.5GHz and consuming 350$\mu$W at a 1.2V supply.

\section*{Acknowledgment}
This work was supported in part by the WiPLASH project founded from the European Union’s Horizon 2020 research and innovation program under Grant Agreement No. 863337.

\end{document}